\pdfoutput=1

\documentclass[11pt]{article}

\usepackage[preprint]{acl}

\usepackage{times}
\usepackage{latexsym}

\usepackage[T1]{fontenc}

\usepackage[utf8]{inputenc}

\usepackage{microtype}
\usepackage{inconsolata}

\usepackage{graphicx}
\usepackage[percent]{overpic}
\usepackage{amsmath}
\usepackage{subcaption}
\usepackage{makecell}
\usepackage{booktabs}
\usepackage{tabularx}
\usepackage{enumitem}
\usepackage{placeins}

\newcolumntype{L}{>{\raggedright\arraybackslash}X}
\newcolumntype{R}{>{\raggedleft\arraybackslash}X}
\newcolumntype{C}[1]{>{\centering\arraybackslash}p{#1}}
\newcolumntype{P}[1]{>{\raggedright\arraybackslash}p{#1}}

\setlist[itemize]{leftmargin=1.05em,labelsep=0.45em,itemsep=0.15em,topsep=0.25em,parsep=0pt,partopsep=0pt}

\title{DiagramBank: Draw from the Great, Draw Your Best}

\author{
  \textbf{Ling Yue\textsuperscript{1}},
  \textbf{Jiaying Wang\textsuperscript{1}},
  \textbf{Zhexi Lu\textsuperscript{1}},
  \textbf{Mingzhi Zhu\textsuperscript{1}},
\\
  \textbf{Tingwen Zhang\textsuperscript{1}},
  \textbf{Zhen Xu\textsuperscript{2}},
  \textbf{Shaowu Pan\textsuperscript{1}\thanks{Corresponding author: \href{mailto:pans2@rpi.edu}{pans2@rpi.edu}.}}
\\
\\
  \textsuperscript{1}Rensselaer Polytechnic Institute,
  \textsuperscript{2}University of Chicago
}

\begin{document}
\maketitle
\raggedbottom
\begin{abstract}
Scientific diagrams help researchers communicate complex methods and relationships, but creating them requires substantial effort in content organization, layout design, and visual refinement.
Automated diagram generation can reduce this workload.
Reference examples can guide this process by showing how scientific ideas are organized and presented visually.
Diagrams from accepted papers provide a rich source of such examples, but their utility as a large-scale reference resource remains underexplored.
We introduce DiagramBank, a diagram--text reference resource containing 57,100 scientific diagrams from 26,751 accepted AI/ML papers.
Each diagram is paired with its caption, paper title, and abstract, with citing paragraphs included when available.
We combine multi-stage filtering, automated checks, and human audits, with an estimated diagram-selection precision of 93.67\%.
Retrieval experiments show that citing paragraphs complement captions with information specific to individual diagrams.
Under a fixed generation pipeline, replacing PaperBanana's curated reference pool with DiagramBank increases the overall PaperBanana score from 4.62 to 7.88 on PaperBananaBench.
Retrieved references also outperform random references on BizGenEval scientific figures.
Additional evaluations with different judges and renderers, together with a separate blind human evaluation, demonstrate the utility of DiagramBank for automated diagram generation.\footnote{\mbox{Code: }\url{https://github.com/csml-rpi/DiagramBank}.}
\end{abstract}

\section{Introduction}
\label{sec:intro}

\begin{figure*}[!t]
    \centering
    \includegraphics[width=\textwidth]{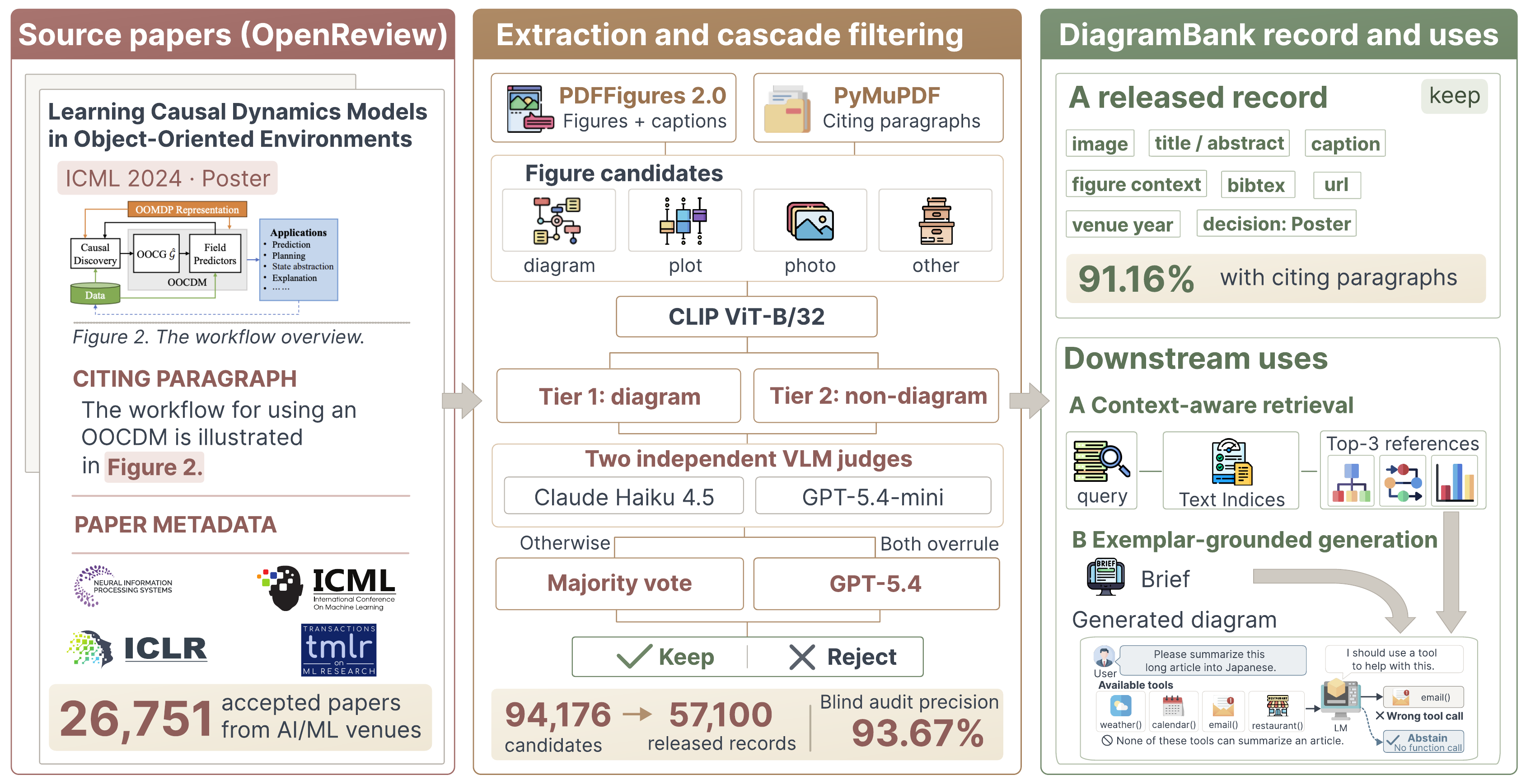}
    \caption{DiagramBank construction and downstream use.
    Left: accepted papers from four OpenReview-hosted AI/ML venues, illustrated by an ICML 2024 diagram with its caption, citing paragraph, and metadata.
    Middle: figure extraction and a CLIP-routed VLM cascade that selects 57,100 records from 94,176 candidates.
    Right: the released record fields and their use in reference retrieval and automated diagram generation.}
    \label{fig:workflow}
\end{figure*}

Scientific diagrams provide compact visual summaries of complex methods, processes, and relationships, helping readers understand the main ideas of a paper~\cite{kembhavi2016diagram}.
However, producing an accurate and clear diagram is labor-intensive.
Researchers must decide which information to include, organize the visual structure, and repeatedly adjust the layout, graphical elements, and text.
These iterations require substantial time and effort before a diagram is ready for publication~\cite{zhu2026paperbananaautomatingacademicillustration,anonymous2026autofigure}.
Automated diagram generation can reduce this manual workload and help researchers communicate their work more efficiently~\cite{mondal2024scidoc2diagrammer}.

Recent diagram generation systems plan diagram content and layout from text and improve their outputs through iterative refinement~\cite{zala2023diagrammergpt,wei2025words2structuredvisuals,anonymous2026autofigure}.
Reference examples further show how scientific concepts can be represented through components, relationships, layouts, and visual styles.
For example, PaperBanana retrieves relevant reference examples to guide content planning and visual style~\cite{zhu2026paperbananaautomatingacademicillustration}.
Retrieving relevant examples requires information about the scientific content they represent.
Their accompanying text provides this information: captions summarize what a diagram shows, while citing paragraphs can explain its components and their relationships~\cite{yang2024scicapplus,subramanian2020medicat,hu2024paperowl}.
Pairing diagrams with explanatory text therefore supports retrieving relevant reference examples, whose visual content can guide new diagram generation.

Existing scientific figure datasets have demonstrated their value in tasks such as captioning, retrieval, and figure reconstruction~\cite{hsu2021scicap,li-etal-2024-multimodal-arxiv,wu-etal-2024-scimmir,rodriguez2023ocrvqgan}.
However, these evaluations do not establish whether large-scale diagram--text resources can serve as reference pools that help systems generate accurate and clear diagrams from scientific text.
Although PaperBanana incorporates reference retrieval into its generation pipeline, it uses a small, manually curated reference pool~\cite{zhu2026paperbananaautomatingacademicillustration}.
This raises an important question: \textit{how can we organize scientific diagrams into a large-scale, reliable reference pool, and how can we demonstrate its utility for automated diagram generation?}

To address this question, we introduce \textbf{DiagramBank}, a dataset designed to make diagrams from scientific papers available as reference examples for diagram generation.
By organizing diagrams together with their explanatory text and provenance, DiagramBank supports the retrieval of examples relevant to a new generation request.
We combine diagram filtering with checks on diagram--text alignment to assess the quality of both the diagrams and their paired text.
We evaluate the resulting resource through diagram retrieval and generation experiments on PaperBananaBench and the scientific-figure domain of BizGenEval~\cite{li2026bizgeneval}.
Our contributions are:
\begin{itemize}
    \item \textbf{A large-scale diagram--text reference resource.} We introduce DiagramBank, a dataset of 57,100 diagrams paired with explanatory text, covering recent AI/ML research. It links diagrams to captions, paper-level context, and citing paragraphs where available, supporting the retrieval of relevant reference examples.
    \item \textbf{A systematic data construction pipeline.} We combine multi-stage filtering, automated checks, and human audits to assess diagram selection and diagram--text alignment. A blind audit estimates 93.67\% diagram-selection precision.
    \item \textbf{Downstream utility.} We show that citing paragraphs provide complementary signals for diagram retrieval. Under a fixed generation pipeline, replacing PaperBanana's curated reference pool with DiagramBank improves the overall PaperBanana score from 4.62 to 7.88 on PaperBananaBench, with gains across all four evaluation dimensions. On BizGenEval scientific figures, retrieved references also outperform random references.
\end{itemize}

\section{Related Work}
\label{sec:related}

\paragraph{Scientific-figure corpora.}
SciCap~\cite{hsu2021scicap} and ArxivCap~\cite{li-etal-2024-multimodal-arxiv} support caption generation at scale, SciMMIR~\cite{wu-etal-2024-scimmir} benchmarks figure--caption and table--caption retrieval, and SPIQA~\cite{pramanick2024spiqa} poses questions about the figures and tables of full papers; DocFigure~\cite{jobin2019docfigure} and ACL-Fig~\cite{karishma2023aclfig} classify figure types.
SciCap+~\cite{yang2024scicapplus}, MedICaT~\cite{subramanian2020medicat}, and M-Paper~\cite{hu2024paperowl} add body text, namely mention paragraphs, inline references, or paragraphs linked through \LaTeX{} references, for graph plots, medical figures, and mixed figures and tables, respectively.
Among diagram-centric resources, AI2D~\cite{kembhavi2016diagram} annotates grade-school science diagrams whose conventions differ from research schematics, and Paper2Fig100k~\cite{rodriguez2023ocrvqgan} collects over 100k architecture and method figures from arXiv with captions, citing paragraphs, and OCR text; visual-summary work ranks representative figures within papers~\cite{yamamoto2021visual,zhong2025smsmo}.
Table~\ref{tab:resource_comparison} compares the closest resources.

\paragraph{Diagram generation and authoring.}
Scientific-diagram generation spans layout planning with language models~\cite{zala2023diagrammergpt,mondal2024scidoc2diagrammer}, graphics-program synthesis trained on TikZ--caption pairs~\cite{belouadi2024automatikz}, text-to-diagram generation and editing benchmarks~\cite{wei2025words2structuredvisuals}, and agentic illustration from long scientific text~\cite{anonymous2026autofigure}.
PaperBanana~\cite{zhu2026paperbananaautomatingacademicillustration} makes retrieval explicit: a retriever selects reference diagrams by their methodology descriptions and captions before planning and rendering, much as retrieval-augmented generation grounds text generation~\cite{lewis2020rag}.
Evaluation sets for diagram authoring hold hundreds to a few thousand curated pairs~\cite{mondal2024scidoc2diagrammer,anonymous2026autofigure,zhu2026paperbananaautomatingacademicillustration}; DiagramBank supplies a larger, audited pool in which every diagram carries the text that such retrievers match against.

\begin{table*}[!t]
\centering
\footnotesize
\setlength{\tabcolsep}{3.5pt}
\renewcommand{\arraystretch}{1.08}
\begin{tabularx}{\textwidth}{@{}P{0.21\textwidth}P{0.115\textwidth}P{0.125\textwidth}P{0.18\textwidth}P{0.17\textwidth}L@{}}
\toprule
\textbf{Resource} & \textbf{Size} & \textbf{Figure types} & \textbf{Text fields} & \textbf{Data curation} & \textbf{Evaluated task} \\
\midrule
\multicolumn{6}{@{}l}{\textit{Scientific figure datasets}} \\
\addlinespace[2pt]
SciCap+\newline\citep{yang2024scicapplus} & 414,809 figures & Graph plots & Caption; first mention paragraph; OCR & Text/OCR check on 200 figures & Captioning \\
\addlinespace[2pt]
MedICaT\newline\citep{subramanian2020medicat} & 217,060 figures & Medical figures & Caption; inline references (74\%) & Subfigure annotations on a subset & Image--text matching \\
\addlinespace[2pt]
ACL-Fig\newline\citep{karishma2023aclfig} & 112,052 figures; 151,900 tables & Mixed figures and tables & Caption; citing sentences (78\%) & 1,671 manually labeled items & Figure classification \\
\addlinespace[2pt]
SciMMIR\newline\citep{wu-etal-2024-scimmir} & 530K figures and tables & Mixed figures and tables & Caption & Hierarchical type annotation & Image--text retrieval \\
\addlinespace[2pt]
M-Paper\newline\citep{hu2024paperowl} & 48,688 papers & Mixed figures and tables & Caption; citing paragraphs & Pairing through \LaTeX{} references & Captioning; analysis \\
\addlinespace[2pt]
Paper2Fig100k\newline\citep{rodriguez2023ocrvqgan} & 102,453 figures & Architectures and pipelines & Caption; citing paragraphs; OCR & Caption and image filters & Figure reconstruction \\
\midrule
\multicolumn{6}{@{}l}{\textit{Diagram generation benchmarks}} \\
\addlinespace[2pt]
SciDoc2\allowbreak{}DiagramBench\newline\citep{mondal2024scidoc2diagrammer} & 1,080 instances & Flowcharts; architectures; summaries; results & Source document; diagram-creation intent & Gold diagrams for a subset & Diagram generation \\
\addlinespace[2pt]
FigureBench\newline\citep{anonymous2026autofigure} & 3,300 pairs & Scientific illustrations & Long-form source text & Human-curated test set; VLM-filtered development set & Diagram generation \\
\addlinespace[2pt]
PaperBananaBench\newline\citep{zhu2026paperbananaautomatingacademicillustration} & 292 test; 292 reference cases & Methodology diagrams & Methodology text; caption & Human screening and annotation & Diagram generation \\
\midrule
\textbf{DiagramBank (ours)} & \textbf{57,100 diagrams} & Scientific diagrams & Caption; citing paragraphs (91.16\%); title; abstract & Multi-stage filtering; automated checks; human audits & Retrieval; diagram generation \\
\bottomrule
\end{tabularx}
\caption{Comparison of scientific figure datasets and diagram generation benchmarks. Size uses the units reported by each source. Data curation summarizes reported filtering, annotation, and quality checks.}
\label{tab:resource_comparison}
\end{table*}

\section{DiagramBank}
\label{sec:diagrambank}

DiagramBank is a diagram--text reference resource for automated diagram generation.
Each record contains a diagram, its caption, and the title and abstract of the source paper.
We also include citing paragraphs when available.
As shown in Figure~\ref{fig:workflow}, we first collect figures and their associated text from scientific papers, then apply a three-stage filtering process to select diagrams.

\subsection{Data Collection}
\label{sec:construction}

We collect accepted papers from four AI/ML venues hosted on OpenReview: ICLR (2017--2026), NeurIPS (2021--2022 and 2024--2025), ICML (2023--2025), and TMLR (2022--2026).
Using the OpenReview Python API~\cite{OpenReviewPythonClient}, we obtain PDFs, paper metadata, and acceptance decisions.
We normalize metadata across venue-specific schemas and identify accepted papers from the recorded decisions.

From each PDF, we extract figures and captions with PDFFigures 2.0~\cite{clark2016pdffigures}, excluding table extractions.
We then use PyMuPDF~\cite{PyMuPDF} to extract body paragraphs associated with each figure by matching figure mentions and figure numbers.
All matching paragraphs are stored as \textit{citing paragraphs}.
Captions summarize what a figure shows, while citing paragraphs can explain its content in more detail.
The title and abstract provide paper-level context for matching reference examples to a generation request.
Each diagram--text record retains the source paper identifier and OpenReview URL for attribution.
Several diagrams can come from the same paper and share its title and abstract.
We preserve their figure-specific captions and citing paragraphs separately so that users can retrieve the diagram describing the required component or procedure.
These source links help exclude target papers from the reference pool used for generation.

\subsection{Data Filtering}
\label{sec:filtering}

We define diagrams as figures whose primary content represents scientific methods, structures, processes, or conceptual relationships through graphical components and their connections.
Examples include architectures, pipelines, flowcharts, computational graphs, and conceptual maps.
The remaining classes are \textit{plot} for quantitative visualizations, \textit{photo} for photographs and related content such as game or simulation screenshots, and \textit{other} for text, equations, tables, or invalid extractions.
Mixed-panel figures are treated as diagrams when diagrammatic content dominates.
We use these four classes throughout the following three stages.

\noindent\textbf{Stage 1: CLIP Routing.}
We use CLIP as a low-cost initial filter to reduce VLM calls while retaining uncertain figures for further review.
OpenCLIP ViT-B/32~\cite{ilharco_gabriel_2021_5143773,radford2021learning} assigns each figure a class and a confidence score based on its similarity to class-specific text prompts.
Figures classified as diagrams at any confidence enter Tier~1; figures assigned another class with confidence below 0.70 enter Tier~2.
These two paths provide 94,176 candidates from accepted papers.
This routing allows the subsequent VLMs to examine both low-confidence diagrams and potentially misclassified figures.

\noindent\textbf{Stage 2: VLM Classification.}
Claude Haiku~4.5 and GPT-5.4-mini independently classify each candidate using the same four-class prompt.
For Tier~1, we retain a figure if at least one VLM labels it as a diagram, giving at least two diagram votes together with CLIP.
If both VLMs reject it, the figure proceeds to Stage~3.
For Tier~2, only figures labeled as diagrams by both VLMs proceed to Stage~3; the remaining candidates are excluded.

\noindent\textbf{Stage 3: Disagreement Resolution.}
GPT-5.4 makes the final decision for candidates where both VLMs disagree with CLIP on whether the figure is a diagram.
These candidates are retained only when GPT-5.4 classifies them as diagrams.
The pipeline retains 57,100 diagrams; Table~\ref{tab:cascade_funnel} summarizes the filtering paths.

We assess diagram selection and diagram--text alignment through automated checks and human audits, with the protocols and results reported in Section~\ref{sec:quality}.
Appendix~\ref{app:cascade_spec} provides the complete prompts, model identifiers, and execution settings.

\begin{table}[t]
\centering
\small
\setlength{\tabcolsep}{5pt}
\renewcommand{\arraystretch}{1.08}
\begin{tabular*}{\columnwidth}{@{\extracolsep{\fill}}lclrr@{}}
\toprule
\textbf{Tier} & \textbf{Votes} & \textbf{Action} & \textbf{Input} & \textbf{Kept} \\
\midrule
1 & 2/2 & Retain & 46{,}524 & 46{,}524 \\
  & 1/2 & Retain & 3{,}645 & 3{,}645 \\
  & 0/2 & Recheck & 10{,}548 & 1{,}865 \\
\addlinespace[4pt]
2 & 2/2 & Recheck & 5{,}172 & 5{,}066 \\
  & 1/2 & Reject & 2{,}602 & 0 \\
  & 0/2 & Reject & 25{,}685 & 0 \\
\midrule
\multicolumn{3}{@{}l}{\textbf{Total}} & \textbf{94{,}176} & \textbf{57{,}100} \\
\bottomrule
\end{tabular*}
\caption{Filtering-path counts from the logs. Votes are diagram labels from the two Stage~2 VLMs; Recheck requires confirmation by GPT-5.4.}
\label{tab:cascade_funnel}
\end{table}

\subsection{Dataset Statistics}
\label{sec:stats}

\begin{figure}[htbp]
\centering
\includegraphics[width=\columnwidth]{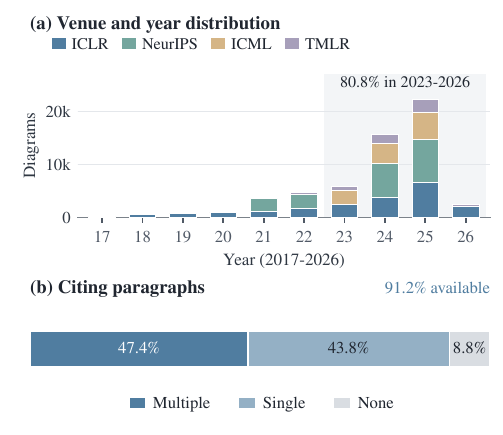}
\caption{Statistics of the 57,100 released diagrams. (a) Counts by venue and year, reflecting release coverage rather than complete annual publication volumes. (b) Shares with multiple, one, or no citing paragraphs.}
\label{fig:compact_corpus_stats}
\end{figure}

DiagramBank contains 57,100 diagrams from 26,751 accepted papers, with 80.84\% of its diagrams from papers dated 2023--2026 (Figure~\ref{fig:compact_corpus_stats}a).
All records include a caption, title, and abstract.
Citing paragraphs are available for 52,052 diagrams (91.16\%), with their coverage shown in Figure~\ref{fig:compact_corpus_stats}b.
Further statistics, record fields, and release details are provided in Appendix~\ref{app:dataset_details}.

\section{Experiments}
\label{sec:utility}

We evaluate DiagramBank from three aspects: the quality of the released diagrams and their paired text, the contribution of the text fields to diagram retrieval, and the utility of the dataset as a reference resource for automated diagram generation.
We keep the generation pipeline fixed while varying the reference input.
This design tests the contribution of the reference resource.

\subsection{Dataset Quality Assessment}
\label{sec:quality}

We assess two aspects of dataset quality: whether the released figures are schematic diagrams and whether their paired text describes them.
Both are important for a reference resource: the images should illustrate scientific methods, and the text should describe the content used to retrieve them.

\noindent\textbf{Diagram quality.}
To evaluate diagram selection, we conduct a blind human audit of 500 released records, with sampling across filtering paths and proportional sampling by venue within each path.
The annotator labels each figure as schematic or non-schematic without seeing CLIP labels, confidence scores, VLM outputs, or filtering paths.
Uncertain filtering paths receive more audit samples relative to their size.
Weighting by the released population prevents these deliberately oversampled paths from dominating the overall precision estimate.
As shown in Table~\ref{tab:release_subsets}, the estimated precision is 93.67\% after weighting each filtering path by its population size.
The 51,590 records labeled as diagrams by both Stage~2 VLMs form \emph{DiagramBank-Core}, with an estimated precision of 95.2\%.
We use this higher-precision subset for the PaperBananaBench reference pool.

The audit also supports retaining uncertain candidates for further review.
The Tier~2 path recovers 5,066 diagrams that CLIP assigns to other classes, with high audited precision after agreement between the two VLMs and confirmation by GPT-5.4.
This finding supports the use of CLIP for initial routing, followed by additional checks that recover diagrams missed by the initial classifier.

\begin{table}[t]
\centering
\small
\setlength{\tabcolsep}{6pt}
\begin{tabular*}{\columnwidth}{@{\extracolsep{\fill}}lrrr@{}}
\toprule
\textbf{Subset} & \textbf{Records} & \textbf{Audited} & \textbf{Precision} \\
\midrule
Full release & 57{,}100 & 500 & 93.7 \\
Core & 51{,}590 & 300 & 95.2 \\
Remaining & 5{,}510 & 200 & 79.3 \\
\bottomrule
\end{tabular*}
\caption{Blind human audit with path-weighted precision (\%). Core and Remaining partition the release. Audit details are in Appendix~\ref{app:audit_protocol}.}
\label{tab:release_subsets}
\end{table}

\noindent\textbf{Diagram--text alignment.}
We separately assess diagram--text alignment in 50 sampled records, including 40 with citing paragraphs.
One author verifies LLM-proposed labels against the images and displayed text excerpts.
Table~\ref{tab:context_audit} shows that all 50 caption--image pairs are broadly aligned and 36 of the 40 citing-paragraph fields explain the diagram.
This review supports the semantic alignment of the sampled pairs; the retrieval experiments below test whether these fields help distinguish diagrams.
See Appendix~\ref{app:text_alignment} for review details.

\begin{table}[t]
\centering
\small
\setlength{\tabcolsep}{7pt}
\begin{tabular*}{\columnwidth}{@{\extracolsep{\fill}}lr@{}}
\toprule
\textbf{Review outcome} & \textbf{Records} \\
\midrule
\multicolumn{2}{@{}l}{\textit{(a) Citing paragraphs}} \\
Explain the diagram & 36 / 40 \\
Limited explanation & 4 / 40 \\
Wrong or uncertain & 0 / 40 \\
\midrule
\multicolumn{2}{@{}l}{\textit{(b) Captions and metadata}} \\
Caption--image pairs broadly aligned & 50 / 50 \\
\quad With minor issues & 3 / 50 \\
Consistent paper metadata & 50 / 50 \\
\bottomrule
\end{tabular*}
\caption{Author-verified review of 50 records, including 40 with citing paragraphs. Appendix~\ref{app:text_alignment} gives the protocol.}
\label{tab:context_audit}
\end{table}

\subsection{Retrieval Evaluation}
\label{sec:retrieval}

\noindent\textbf{Setup.}
We conduct a controlled retrieval experiment to assess how different text fields help recover a target diagram from DiagramBank.
The candidate pool contains 52,052 released records with a caption, title, abstract, and citing paragraphs.
We sample 2,000 target diagrams stratified by venue, year, and caption length, selecting at most one diagram per paper.
For each target, a language model generates a 35--90-word query from its four text fields, and an automatic validator rejects queries containing source-identifying names or copied text spans.
We use text-embedding-3-large~\cite{OpenAIEmbeddings2024} to encode each query and each candidate text field separately.
A retrieval script ranks all candidates by the mean cosine similarity between the query and the selected fields.
All conditions use the same queries, candidate pool, and embedding model; only the candidate fields used for scoring change.
We report R@1, R@10, and MRR@10, treating the original diagram associated with each query as the retrieval target.
The full protocol is provided in Appendix~\ref{app:retrieval_protocols}.

\begin{table}[t]
\centering
\footnotesize
\setlength{\tabcolsep}{3pt}
\begin{tabularx}{\columnwidth}{@{}Lrrr@{}}
\toprule
\textbf{Candidate fields} & \textbf{R@1} & \textbf{R@10} & \textbf{MRR@10} \\
\midrule
\multicolumn{4}{@{}l}{\textit{(a) Field comparison: 2,000 queries}} \\
Caption only & \textbf{72.90} & 86.60 & 77.69 \\
Title, abstract, caption & 64.95 & 85.75 & 72.00 \\
Full (all four fields) & 68.60 & 88.95 & 75.79 \\
Full, wrong paragraphs & 56.60 & 80.80 & 64.56 \\
Full without title & 71.75 & \textbf{90.75} & \textbf{78.48} \\
\midrule
\multicolumn{4}{@{}l}{\textit{(b) Same-paper paragraph swap: 913 queries}} \\
Caption only & \textbf{72.51} & 85.98 & \textbf{77.27} \\
Full without paragraphs & 61.99 & 84.67 & 69.84 \\
Full (all four fields) & 66.48 & \textbf{88.61} & 73.98 \\
Full, same-paper paragraph swap & 49.73 & 82.15 & 61.33 \\
\bottomrule
\end{tabularx}
\caption{Exact-target retrieval (\%). Both panels use 52,052 candidates; panel (b) uses the 913 queries eligible for a same-paper paragraph swap. Bold marks the best score per panel. See Table~\ref{tab:retrieval_full} for full ablations.}
\label{tab:retrieval_ablation}
\label{tab:within_paper_context}
\end{table}

\noindent\textbf{Results.}
Table~\ref{tab:retrieval_ablation} shows that adding citing paragraphs to the title, abstract, and caption increases R@10 from 85.75 to 88.95.
The full field set also outperforms captions alone on this metric.
Citing paragraphs thus provide complementary information that helps retrieve target diagrams missed by captions and paper-level text.
Captions provide a strong starting point for retrieval, while citing paragraphs improve coverage among the top ten candidates.
This is useful when a generation system selects several reference examples for a request.
The best R@10 uses the caption, abstract, and citing paragraphs, combining figure descriptions with explanations of their scientific content.

This benefit depends on pairing each diagram with the correct paragraphs.
For the first control, we replace citing paragraphs with passages of similar length from other papers of the same venue and year.
Retrieval performance decreases even though the candidates still have the same text fields and comparable paragraph lengths.
To separate figure-specific information from paper-level topic information, we also swap citing paragraphs between different diagrams of the same paper.
This swap reduces R@10 by 6.46 points on the 913 eligible queries.
The query, image, caption, title, and abstract remain unchanged; only the paragraph assigned to the candidate changes.
This result shows that citing paragraphs carry information specific to individual diagrams, beyond the title and abstract shared within a paper.

\subsection{Diagram Generation}
\label{sec:generation}

\noindent\textbf{Setup.}
We evaluate DiagramBank as a reference resource for automated scientific diagram generation using the PaperBanana pipeline~\cite{zhu2026paperbananaautomatingacademicillustration} on all 292 PaperBananaBench test cases.
We keep the generation pipeline, models, prompt templates, retriever, and reranker fixed, and vary the reference pool.
Qwen3.8-27B (Qwen3.8) serves as planner and stylist, and Qwen-Image-2.1 serves as renderer.
We compare no references with ten references from each of three pools: PaperBanana's original pool of 292 curated diagrams, Paper2Fig100k~\cite{rodriguez2023ocrvqgan}, and DiagramBank.
The no-reference condition measures the benefit of adding examples, while the existing reference pools test whether the choice of corpus matters.
Using ten references for each pool keeps the number of examples fixed across these comparisons.
The ten selected references include at most one diagram per paper.
DiagramBank references are drawn from Core records with citing paragraphs, with paper, author, and text-overlap exclusions to reduce leakage.
Each DiagramBank reference contains its image, abstract, and citing paragraphs, providing visual examples and scientific explanations for the planner.
The generator receives the target description and references from other papers when applicable; the target diagram from the original paper is used only for evaluation.
Appendix~\ref{app:generation_protocols} provides the implementation details and reference exclusion rules.

\begin{figure*}[t]
\centering
\begin{subfigure}[t]{\textwidth}
    \centering
    \includegraphics[trim=0bp 340bp 0bp 160bp,clip,width=\linewidth]{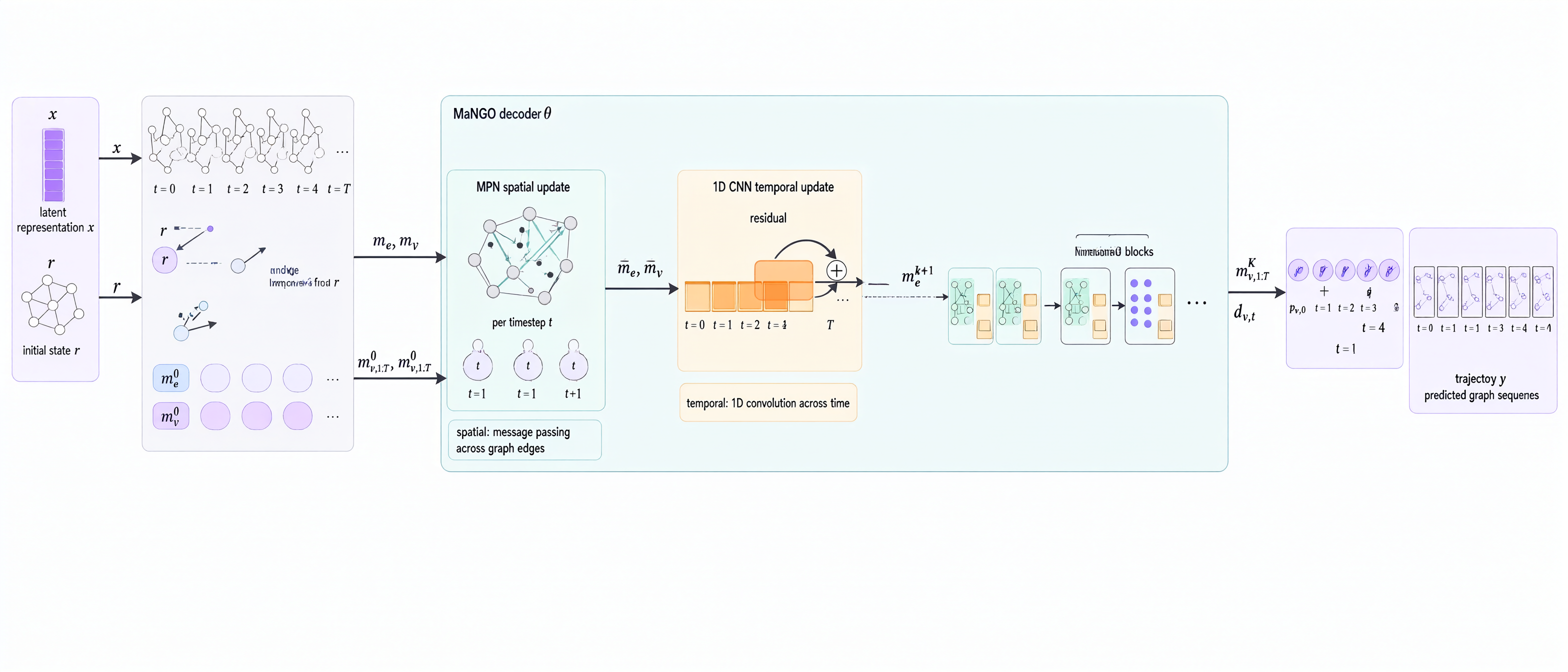}
    \caption{No references}
    \label{fig:pbb_example_c0}
\end{subfigure}

\begin{subfigure}[t]{\textwidth}
    \centering
    \includegraphics[trim=0bp 14bp 0bp 20bp,clip,width=\linewidth]{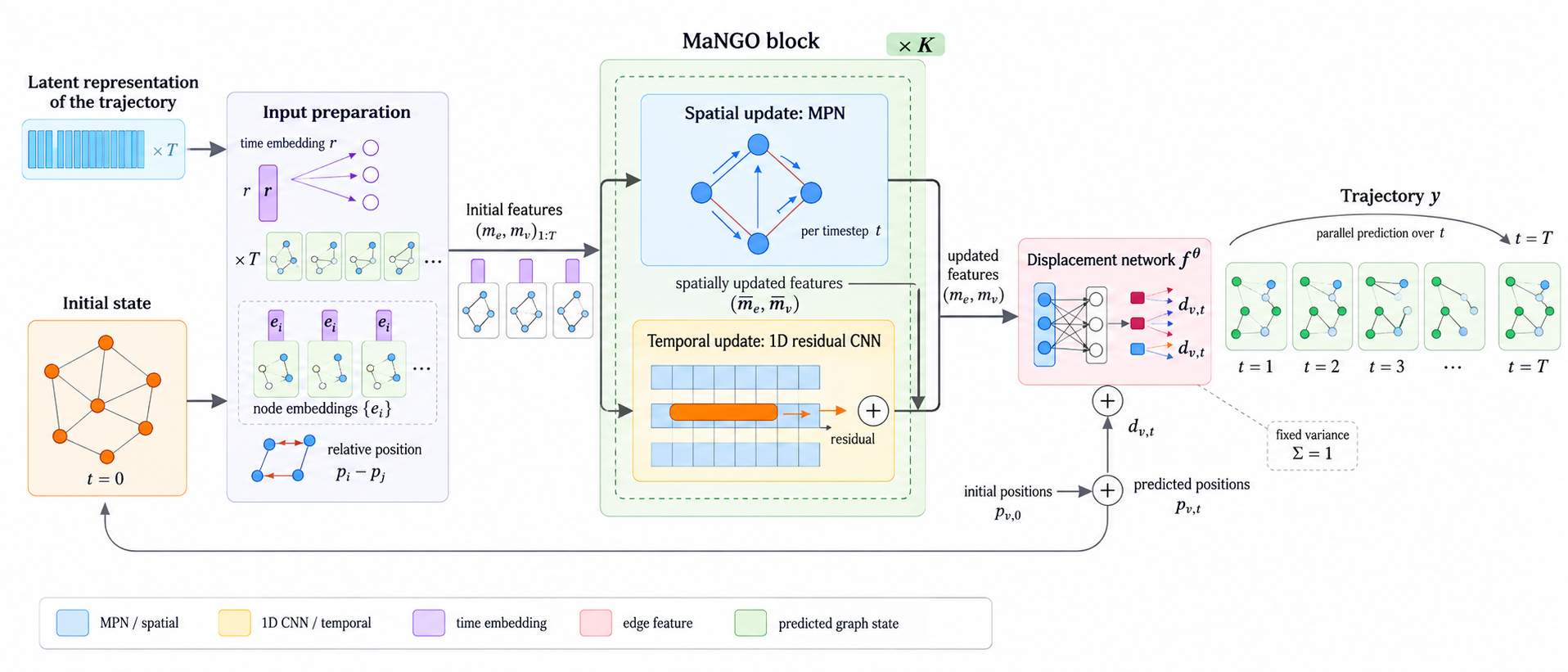}
    \caption{DiagramBank references}
    \label{fig:pbb_example_c2}
\end{subfigure}
\caption{Generation with and without DiagramBank references for the MaNGO trajectory-decoder case in PaperBananaBench. Both outputs use the same planner and renderer. DiagramBank references improve label readability and the organization of the spatial--temporal block and displacement network.}
\label{fig:pbb_example}
\end{figure*}

\noindent\textbf{Evaluation.}
We use Qwen3.8 as the judge for two complementary evaluations.
For the PaperBanana score, a win against the target diagram receives 100, a tie 50, and a loss 0.
It measures faithfulness, conciseness, readability, and aesthetics.
The overall verdict prioritizes faithfulness and readability, then uses conciseness and aesthetics to resolve the remaining cases.
A separate pairwise rubric compares outputs from two generation conditions for faithfulness, text, structure, and visual quality.
This evaluation tests whether changing the reference input improves the output for the same generation request.
The judge sees the target diagram and evaluates both presentation orders.
Pairwise preference scores range from $-1$ to $1$; positive scores favor the first condition.
We also report wins, ties, and losses.

\begin{table}[t]
\centering
\small
\setlength{\tabcolsep}{4pt}
\begin{tabular*}{\columnwidth}{@{\extracolsep{\fill}}lrrrr@{}}
\toprule
& \multicolumn{2}{c}{\textbf{Score}} & \multicolumn{2}{c@{}}{\textbf{Gain over}} \\
\cmidrule(lr){2-3}\cmidrule(l){4-5}
\textbf{Metric} & \textbf{None} & \textbf{DB} & \textbf{None} & \textbf{PB} \\
\midrule
Faithfulness & 6.68 & \textbf{10.10} & +3.42 & +2.05 \\
Conciseness & 6.85 & \textbf{7.53} & +0.68 & +3.08 \\
Readability & 9.42 & \textbf{12.67} & +3.25 & +2.23 \\
Aesthetics & 11.47 & \textbf{15.24} & +3.77 & +1.88 \\
\midrule
Overall & 3.77 & \textbf{7.88} & +4.11 & +3.25 \\
\bottomrule
\end{tabular*}
\caption{PaperBanana scores on 292 cases (0--100), judged by Qwen3.8 against the target diagram. DB: DiagramBank; PB: PaperBanana references (overall 4.62); None: no references. Each pool supplies ten references. Higher is better; gains use unrounded scores.}
\label{tab:pbb}
\end{table}

\begin{table}[t]
\centering
\small
\setlength{\tabcolsep}{4pt}
\begin{tabular*}{\columnwidth}{@{\extracolsep{\fill}}lrr@{}}
\toprule
\textbf{Comparison} & \textbf{Preference} & \textbf{W/T/L} \\
\midrule
DB vs.\ PB & +0.188 & 151/45/96 \\
DB vs.\ None & +0.404 & 190/30/72 \\
Paper2Fig100k vs.\ None & +0.18 & 156/32/104 \\
\bottomrule
\end{tabular*}
\caption{Pairwise preference scores on the same 292 cases, judged by Qwen3.8. Positive scores in $[-1,1]$ favor the first condition; W/T/L counts its wins, ties, and losses. DB: DiagramBank; PB: PaperBanana; None: no references. Appendix Table~\ref{tab:pbb_rubric_details} gives the full criterion scores and Haiku~4.5 results.}
\label{tab:pbb_rubric}
\end{table}

\noindent\textbf{Results on PaperBananaBench.}
Table~\ref{tab:pbb} shows that replacing PaperBanana's curated reference pool with DiagramBank improves the overall PaperBanana score from 4.62 to 7.88, with gains across all four dimensions.
DiagramBank also improves generation over no references.
Direct pairwise evaluation in Table~\ref{tab:pbb_rubric} also favors DiagramBank, with 151 wins against 96 losses compared with PaperBanana references over the same 292 cases.
Both DiagramBank and Paper2Fig100k improve on the no-reference baseline; DiagramBank receives the higher pairwise preference score under the same retrieval and evaluation setup.
The improvements cover all four criteria, with the largest gains over PaperBanana references in structure and visual quality.
This pattern is consistent with using published diagrams as examples of component organization and layout.
Together, these results demonstrate the utility of DiagramBank as a reference resource for automated scientific diagram generation.

Figure~\ref{fig:pbb_example} illustrates the improvements in structure and visual quality for the MaNGO trajectory-decoder case.
The output without references includes the main processing modules but contains garbled labels and repeated time indices.
With DiagramBank references, the output connects input preparation, the repeated spatial--temporal block, the displacement network, and the predicted trajectory in a clear sequence.
Readable labels and a color legend explain the organization of these components.
These changes illustrate how reference examples can help present the relationships between scientific components.

Claude Haiku~4.5 also prefers DiagramBank outputs to those generated without references.
Both Qwen3.8 and Haiku~4.5 give positive scores on all four criteria across the 292 cases.
On a 48-case subset, Claude Sonnet~5.5 and Qwen3.8 assign DiagramBank the same overall pairwise preference score.
We then reuse each condition's rendering prompts with Sonnet~5.5 as an SVG renderer in place of Qwen-Image-2.1.
The reference condition improves the overall PaperBanana score with both renderers.
These additional results support the use of DiagramBank references for both image generation and code-based diagram generation.
Appendix~\ref{app:generation_protocols} reports the comparisons, and Figure~\ref{fig:svg_example} shows an SVG example.

\noindent\textbf{Results on BizGenEval.}
We use BizGenEval~\cite{li2026bizgeneval} to test whether retrieving relevant examples improves generation over random reference selection.
Its scientific-figure domain contains 80 prompts evaluated by Qwen3.8 using the benchmark's checklists.
Each prompt is evaluated using 20 binary checklist questions.
We report question accuracy and the benchmark's Easy, Hard, and All scores.
We compare no references, ten random DiagramBank references, and ten retrieved DiagramBank references under the same generation pipeline.
Table~\ref{tab:bizgeneval} shows that retrieved references perform best on all four metrics, raising the All score from 53.4 with random references to 55.6.
These results support selecting references that match the scientific content of the generation request.
Because random and retrieved references use the same number of images, this comparison tests the contribution of reference selection within DiagramBank.
Appendix~\ref{app:bizgeneval} reports the full protocol and results for other domains.

\begin{table}[t]
\centering
\small
\setlength{\tabcolsep}{5pt}
\begin{tabular*}{\columnwidth}{@{\extracolsep{\fill}}lrrrr@{}}
\toprule
\textbf{References} & \textbf{Acc.} & \textbf{Easy} & \textbf{Hard} & \textbf{All} \\
\midrule
None & 73.1 & 67.0 & 42.0 & 51.1 \\
Random & 73.3 & 69.8 & 40.8 & 53.4 \\
Retrieved & \textbf{75.1} & \textbf{72.5} & \textbf{42.5} & \textbf{55.6} \\
\bottomrule
\end{tabular*}
\caption{BizGenEval scientific-figure results on 80 prompts, using no references, ten random DiagramBank references, or ten retrieved ones. Higher is better.}
\label{tab:bizgeneval}
\end{table}

\noindent\textbf{Blind human evaluation.}
We construct 20 scientific briefs from DiagramBank records, five per venue.
Codex's image generation tool renders each brief once with three retrieved references and once without references.
Both conditions use identical text prompts, and retrieval excludes the target paper.
Four evaluators independently compare the outputs for faithfulness, readable labels, and visual organization, with randomized case order and counterbalanced left/right positions.
They see the scientific brief and both generated diagrams, while reference conditions and source identifiers are hidden.
A condition wins a case when at least three of four evaluators prefer it; all other patterns count as ties.
Table~\ref{tab:human_summary} shows that DiagramBank wins 14 of the 20 cases, while the condition without references wins one.
Each of the four evaluators prefers DiagramBank in more than half of the cases.
These human preferences support the improvements observed in automated evaluation.
Appendix~\ref{app:human_evaluation} gives the full protocol.

\begin{table}[t]
\centering
\small
\setlength{\tabcolsep}{6pt}
\begin{tabular*}{\columnwidth}{@{\extracolsep{\fill}}lrrr@{}}
\toprule
\textbf{Unit} & \textbf{DB} & \textbf{None} & \textbf{Tie} \\
\midrule
Case outcomes (20) & \textbf{14} & 1 & 5 \\
Individual votes (80) & \textbf{59} & 14 & 7 \\
\bottomrule
\end{tabular*}
\caption{Blind human evaluation of 20 diagram pairs. DB: DiagramBank references; None: no references. Four evaluators judge each pair; a case win requires at least three votes.}
\label{tab:human_summary}
\end{table}

\section{Conclusion}
\label{sec:conclusion}
DiagramBank provides 57,100 scientific diagrams from accepted AI/ML papers as a diagram--text reference resource.
It links these diagrams to explanatory text and source papers, making published examples available for retrieval and use in generation systems.
Its construction combines multi-stage filtering, automated checks, and human audits.
Retrieval experiments show that correctly paired citing paragraphs complement captions in locating target diagrams.
Generation experiments show improvements over both no references and PaperBanana's curated pool under fixed pipelines.
On BizGenEval, retrieved references outperform random references.
Blind human evaluation supports these gains and the utility of DiagramBank as a reference resource for automated diagram generation.

\section*{Limitations}
\label{sec:limitations}

\noindent\textbf{Coverage.}
DiagramBank focuses on schematic diagrams from accepted AI/ML papers at four OpenReview-hosted venues.
Its coverage reflects the venues and years included in the collection, with most diagrams coming from recent papers.
Scientific fields with different visual conventions may require additional data and adapted filtering criteria.

\noindent\textbf{Evaluation scope.}
Our evaluation covers controlled retrieval and scientific diagram generation with the tested pipelines.
Additional model judges and a separate blind human evaluation provide complementary evidence.
Future work can extend the evaluation to user-written retrieval queries and repeated generation runs.

\section*{Ethical Considerations}
\label{sec:ethics}

DiagramBank is derived from public scholarly documents and can be used in authoring pipelines, so provenance, redistribution, privacy, bias, misuse, and AI-use disclosure matter.
Figures originate from third-party papers and remain subject to the applicable source terms; the release preserves source identifiers and bibliographic entries for attribution.
The source-document collection does not intentionally target non-public personal data, but publication metadata and figure contents may still contain identifying information from source papers; records include public author names and OpenReview identifiers, decisions, and, where OpenReview exposes them, reviewer scores, which we keep for provenance and corpus analysis, and the intended-use statement excludes author profiling.
The corpus reflects OpenReview-era AI/ML publication practices and should not be treated as demographically or scientifically representative.
Generated figures using retrieved references can be misleading if used without disclosure or checking; users should track provenance, cite both DiagramBank and the original source papers when specific diagrams inform downstream work, disclose AI assistance where appropriate, manually verify factual and visual details, and honor reasonable rights-holder takedown requests~\cite{klug2024can}.
\paragraph{Use of generative AI.}
In this work, we used generative AI tools to implement and execute parts of the experimental pipeline, namely to write and debug the scripts for figure extraction, filtering, retrieval, diagram generation, and evaluation, and to launch and monitor these runs.
Generative models are also components of the documented methods: Claude Haiku~4.5, GPT-5.4-mini, and GPT-5.4 classify candidate figures in the filtering cascade (Section~\ref{sec:filtering}); \texttt{gpt-5.6-sol} generates the retrieval queries used in the ablation study (Appendix~\ref{app:retrieval_protocols}); in the generation experiments, Qwen3.8-27B plans diagrams and reranks references, Qwen-Image-2.1, Claude Sonnet~5.5, and Codex\textquotesingle s \texttt{image\_gen} tool render them, and Qwen3.8-27B, Claude Haiku~4.5, and Claude Sonnet~5.5 judge them (Section~\ref{sec:generation}); and LLM agents proposed the labels of the 50-record diagram--text review, which an author then checked (Appendix~\ref{app:text_alignment}).
Figures~\ref{fig:pbb_example} and~\ref{fig:svg_example} therefore show AI-generated diagrams, whereas the authors drew Figure~\ref{fig:workflow} in Figma.
We have not used generative AI tools to develop conceptual frameworks, propose or refine hypotheses, or design research methodology or experiments.
Additionally, we used generative AI tools to edit the paper for readability, to create and debug software code, and to search for and identify relevant literature.
We have reviewed all AI-assisted work: AI-edited text was checked by the authors for meaning and accuracy; AI-assisted code was reviewed and tested by the authors, and the experiment runs were verified against their logs and outputs; and all literature searched by AI tools was read by the authors before being cited.
We take responsibility for the final content of this work, including text, claims, or artifacts produced with the aid of generative AI.

\bibliography{main}

\clearpage
\appendix

\section{Dataset Details}
\label{app:dataset_details}

\subsection{Dataset Statistics}
\label{app:stat_charts}

\begin{figure*}[t]
\centering
\includegraphics[width=\textwidth]{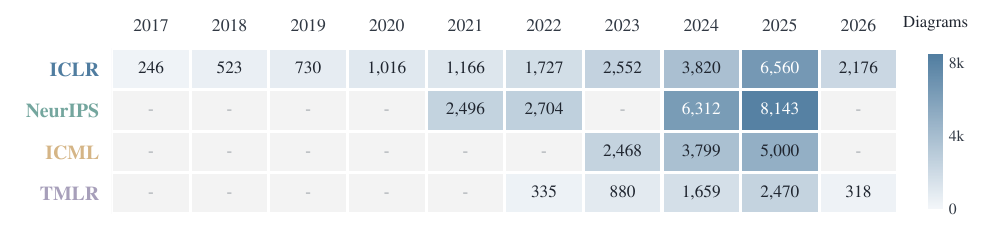}
\caption{Released diagrams by venue and year. Each cell gives the number of diagrams, totaling 57,100. A dash marks a venue and year with no records in this release. Shading scales with the diagram count; venue label colors match Figure~\ref{fig:compact_corpus_stats}.}
\label{fig:dataset_stats}
\end{figure*}

Table~\ref{tab:venue_statistics} summarizes the 57,100 released diagrams and their source papers.
Paper counts refer to distinct OpenReview identifiers, and caption lengths count whitespace-separated words.
Figure~\ref{fig:dataset_stats} gives the complete venue and year distribution.
These counts describe the released dataset under the collection coverage in Section~\ref{sec:construction}.

\begin{table}[htbp]
\centering
\small
\setlength{\tabcolsep}{6pt}
\begin{tabular*}{\columnwidth}{@{\extracolsep{\fill}}lrrr@{}}
\toprule
\textbf{Venue} & \textbf{Diagrams} & \textbf{Papers} & \makecell{\textbf{Caption}\\\textbf{words}} \\
\midrule
ICLR & 20{,}516 & 9{,}466 & 47.20 \\
NeurIPS & 19{,}655 & 9{,}802 & 49.96 \\
ICML & 11{,}267 & 5{,}148 & 49.14 \\
TMLR & 5{,}662 & 2{,}335 & 52.91 \\
\midrule
Total & 57{,}100 & 26{,}751 & 49.10 \\
\bottomrule
\end{tabular*}
\caption{Dataset statistics by venue. Caption words reports the mean length per diagram; the total is weighted by the number of diagrams.}
\label{tab:venue_statistics}
\end{table}

\subsection{Record Fields and Paper Groups}
\label{app:full_schema}

Each record is identified by its paper identifier and figure number.
Table~\ref{tab:release_schema} groups the released fields by their role in retrieval, attribution, and quality control.
The diagram image is linked through a relative file path.
Captions, titles, and abstracts are present for all records; citing paragraphs are available for 52,052 records.
The database stores citing paragraphs in \texttt{figure\_context}; we use \emph{citing paragraphs} throughout the paper to refer to this field.

\begin{table*}[t]
\centering
\small
\setlength{\tabcolsep}{5pt}
\renewcommand{\arraystretch}{1.10}
\begin{tabularx}{\textwidth}{@{}P{0.15\textwidth}P{0.39\textwidth}L@{}}
\toprule
\textbf{Group} & \textbf{Fields} & \textbf{Description} \\
\midrule
Source identifiers & \texttt{platform}, \texttt{platform\_id}, \texttt{url}, \texttt{bibtex} & OpenReview source identifiers, source URL, and bibliographic entry for attribution. \\
Paper text & \texttt{title}, \texttt{abstract}, \texttt{tldr} & Paper title and abstract, with an optional TL;DR summary. \\
Paper metadata & \texttt{venue}, \texttt{year}, \texttt{authors}, \texttt{author\_ids}, \texttt{keywords}, \texttt{areas}, \texttt{decision}, \texttt{scores}, \texttt{cdate} & Venue, year, authors, optional subject and review metadata, and the source-record creation date. \\
Diagram and text & \texttt{figure\_number}, \texttt{figure\_path}, \texttt{figure\_caption}, \texttt{figure\_context} & Figure identifier, image path, caption, and citing paragraphs. \\
CLIP outputs & \texttt{clip\_type}, \texttt{clip\_confidence} & Predicted figure class and routing confidence. The database names these fields \texttt{figure\_type} and \texttt{confidence}. \\
VLM outputs & \texttt{label\_haiku}, \texttt{label\_gpt\_mini}, \texttt{label\_gpt\_tiebreak} & Saved labels from the two Stage~2 VLMs and GPT-5.4. The last field is null when confirmation was not requested. \\
Filtering decisions & \texttt{cascade\_path}, \texttt{label\_cascade} & Retained filtering path and final decision, supporting selection of DiagramBank-Core and other subsets. \\
\bottomrule
\end{tabularx}
\caption{Released record fields. Paper and diagram fields are stored in the DuckDB \texttt{Data} table; filtering outputs are also provided in \texttt{cascade\_labels.jsonl}. Optional metadata can be missing. BibTeX entries are unavailable for the ICML 2023 records.}
\label{tab:release_schema}
\end{table*}

\noindent\textbf{Paper groups.}
We group paper records that have identical nonempty titles and abstracts after Unicode case folding and whitespace normalization, linking 312 paper identifiers into 156 pairs.
These pairs are from NeurIPS 2021 and cover 562 released diagram records.
The groups identify papers listed under multiple identifiers.
The file \texttt{paper\_groups.jsonl} records these groups.
All diagrams from a paper and all identifiers in the same group should remain together when constructing data splits.
The generation studies also exclude records from the target paper as described in Appendix~\ref{app:generation_protocols}.

\subsection{Access, Attribution, and Intended Use}
\label{sec:dataset_card}

DiagramBank is distributed through the resource link in the abstract.
The release contains image archives, a DuckDB database~\cite{raasveldt2019duckdb}, FAISS indices~\cite{douze2025faiss}, and filtering labels, with source identifiers retained for attribution.
We document the dataset's scope, record fields, access, and intended use following datasheet and data-statement practices~\cite{gebru2021datasheets,bender-friedman-2018-data}.

The dataset is intended for scientific diagram retrieval and as a source of reference examples for diagram generation.
Its unit is a complete diagram paired with text and paper metadata.
It does not include component-level bounding boxes, node and edge annotations, or a fixed train/test split.
Users can form task-specific subsets using text availability, venue metadata, and filtering paths.

The code is distributed under the MIT License.
Figure rights remain with the source papers, and use of the original figures is subject to the applicable source terms.
All 57,100 records contain a source URL, and 54,632 contain a BibTeX entry.
Users should attribute specific diagrams to their source papers as well as cite DiagramBank, and the maintainers will honor reasonable rights-holder takedown requests.

\section{Construction and Quality Assessment}
\label{app:construction_quality}

\subsection{Filtering Prompts and Configuration}
\label{app:cascade_spec}

Table~\ref{tab:filtering_prompts} provides the class descriptions used by CLIP and the three VLMs.
For CLIP, we average the text embeddings of the phrases for each class to obtain a class centroid.
Classification uses the similarity between the normalized image embedding and the four centroids, with a softmax over $100\times$ cosine similarity.
The routing and voting rules are specified in Section~\ref{sec:filtering}.

\begin{table*}[t]
\centering
\small
\setlength{\tabcolsep}{5pt}
\renewcommand{\arraystretch}{1.10}
\begin{tabularx}{\textwidth}{@{}P{0.075\textwidth}LL@{}}
\toprule
\textbf{Class} & \textbf{CLIP prompt phrases} & \textbf{Shared VLM class description} \\
\midrule
diagram & a scientific diagram or schematic; a system architecture or flow chart; a neural network or computational graph; a block diagram with arrows & schematic, flowchart, system architecture, pipeline, block diagram, neural network graph, or computational graph \\
plot & a statistical data chart; a line chart or scatter plot with axes; a bar chart or histogram; a confusion matrix heatmap; a quantitative visualization of results; curves showing performance metrics & statistical chart, line/scatter/bar plot, histogram, heatmap, confusion matrix, or any quantitative data visualization \\
photo & a photograph of a real object; a microscopy or camera image; a realistic photo without text; a natural scene; a screenshot of a video game or simulation; a pixel art illustration or comic strip & natural photograph, camera/microscopy image, screenshot of a video game or simulation, pixel art, or comic-style illustration \\
other & a full page of dense text; a standalone mathematical formula; a table of numbers without graphics; garbage noise or corruption; a blank white page & dense text, standalone math equations, tables of numbers without graphics, blank pages, or anything that does not fit above \\
\bottomrule
\end{tabularx}
\caption{Verbatim CLIP prompt phrases and VLM class descriptions used in the filtering cascade. The VLM task and output instructions are given in the text.}
\label{tab:filtering_prompts}
\end{table*}

The shared VLM prompt begins with ``Classify this scientific figure into exactly one category: diagram, plot, photo, or other.''
After the class descriptions, it ends with ``Reply with ONLY the category name (one word).''
The GPT models also receive the instruction ``You are a scientific figure classifier.''
Responses are trimmed and lowercased, and the first matching class substring is selected in the order diagram, plot, photo, other.
Unmatched responses are preserved; only the stored label \texttt{diagram} counts as a positive vote.

Table~\ref{tab:model_settings} gives the model settings.
The saved votes reproduce the 57,100 retained records and path totals in Table~\ref{tab:cascade_funnel}.

\begin{table*}[t]
\centering
\small
\setlength{\tabcolsep}{5pt}
\renewcommand{\arraystretch}{1.10}
\begin{tabularx}{\textwidth}{@{}P{0.14\textwidth}P{0.43\textwidth}L@{}}
\toprule
\textbf{Stage} & \textbf{Model} & \textbf{Execution settings} \\
\midrule
Stage 1 & OpenCLIP ViT-B/32 & Four class centroids; normalized embeddings; batch size 16. \\
Stage 2 & Claude Haiku 4.5 & Temperature 0; maximum 16 output tokens; five attempts; eight workers. \\
Stage 2 & \texttt{gpt-5.4-mini} & PNG input; default decoding; streaming; five attempts; eight workers. \\
Stage 3 & \texttt{gpt-5.4} & Same prompt and input format; default decoding; streaming; five attempts; four workers. \\
\bottomrule
\end{tabularx}
\caption{Model settings of the filtering cascade. The GPT calls used the endpoint's default decoding settings and \texttt{store=False}.}
\label{tab:model_settings}
\end{table*}

The reproduction environment specifies \texttt{open\_clip\_torch} 3.2.0, \texttt{torch} 2.9.1, PyMuPDF 1.26.7, \texttt{openreview-py} 1.54.5, DuckDB 1.4.1, and \texttt{faiss-cpu} 1.10.0.
Figures and captions are extracted with PDFFigures 2.0.

\subsection{Audit of Released Diagrams}
\label{app:audit_protocol}

The diagram-quality audit samples 500 records from the final release, with higher sampling rates for uncertain filtering paths and proportional sampling by venue within each path.
One human annotator labels each image as schematic or non-schematic without seeing its filtering path, CLIP predictions, confidence, or VLM labels.
Table~\ref{tab:release_audit_paths} reports the audit counts for each filtering path.

\begin{table}[t]
\centering
\footnotesize
\setlength{\tabcolsep}{3pt}
\begin{tabularx}{\columnwidth}{@{}Xrrrr@{}}
\toprule
\textbf{Path} & $N_h$ & $n_h$ & $\mathrm{TP}_h$ & \textbf{Prec.} \\
\midrule
Tier 1 unanimous & 46{,}524 & 100 & 95 & 95.0 \\
Tier 1 majority & 3{,}645 & 75 & 62 & 82.7 \\
Tier 1 minority + GPT & 1{,}865 & 125 & 91 & 72.8 \\
Tier 2 consensus + GPT & 5{,}066 & 200 & 194 & 97.0 \\
\midrule
Total & 57{,}100 & 500 & 442 & 93.7 \\
\bottomrule
\end{tabularx}
\caption{Audit counts by filtering path. $N_h$ is the path population, $n_h$ the sample size, and $\mathrm{TP}_h$ the number labeled schematic. Precision is a percentage; the total uses population weighting. GPT denotes GPT-5.4 confirmation.}
\label{tab:release_audit_paths}
\end{table}

For path $h$, let $W_h=N_h/57{,}100$ and $\hat p_h=\mathrm{TP}_h/n_h$.
We estimate release precision as
\begin{equation}
\hat p = \sum_h W_h\hat p_h.
\end{equation}
This gives 93.67\% precision.
For DiagramBank-Core and the remaining records, the same calculation normalizes the population weights within each subset (Table~\ref{tab:release_subsets}).
Screenshots and photo-like figures account for the main observed errors, particularly on the Tier~1 minority path.

\subsection{Diagram--Text Alignment}
\label{app:text_alignment}

\noindent\textbf{Automated checks.}
Table~\ref{tab:context_automatic_checks} reports text availability and metadata consistency over the full release.
The extraction rule retains a paragraph when ``Figure'' or ``Fig.'' is followed within 50 characters by the stored figure number.
Among the 52,052 records with citing paragraphs, 51,321 contain an explicit reference to the target figure number as a single reference, list, or range.
The remaining 731 are linked by proximity to a figure mention under the extraction rule.
These checks describe field availability and reference matching; the review below assesses semantic alignment.

\begin{table}[t]
\centering
\small
\setlength{\tabcolsep}{4pt}
\begin{tabularx}{\columnwidth}{@{}Lr@{}}
\toprule
\textbf{Check} & \textbf{Records} \\
\midrule
Title, abstract, caption, and URL present & 57{,}100 / 57{,}100 \\
Citing paragraphs available & 52{,}052 / 57{,}100 \\
Explicit target-figure reference among records with citing paragraphs & 51{,}321 / 52{,}052 \\
Consistent metadata within each paper & 26{,}751 / 26{,}751 \\
\bottomrule
\end{tabularx}
\caption{Release-wide text and metadata checks. The last row counts papers; consistency requires identical title, abstract, URL, venue, and year across their records.}
\label{tab:context_automatic_checks}
\end{table}

\noindent\textbf{Author-verified review.}
A separate deterministic sample of 50 records (seed 20260709) contains 40 with citing paragraphs and 10 without.
The 40 records comprise 14 ICLR, 11 ICML, 14 NeurIPS, and one TMLR example.
Each review shows the image, caption, citing paragraphs when available, figure number, title, and abstract.
LLMs propose initial labels, and one author verifies them against the displayed content.

Citing paragraphs are labeled as explaining the diagram, referring to it with limited explanation, or being wrong or uncertain.
Caption--image pairs are labeled as a clear match, a broad match with minor issues, or a mismatch.
The review displays up to 500 characters per caption and 900 characters per citing-paragraph field; labels assess these displayed excerpts.
Table~\ref{tab:context_audit} reports the results: 36 of 40 paragraph fields explain their diagram, and all 50 caption--image pairs are broadly aligned, including three with minor issues.
These analyses use the stored text of the evaluated release.

\FloatBarrier

\section{Retrieval Evaluation}
\label{app:retrieval_protocols}

\subsection{Candidate Pool and Query Construction}

The candidate pool contains 52,052 records from 25,616 papers with nonempty captions, titles, abstracts, and citing paragraphs.
Each retrieval target is identified by its paper identifier and figure number.
We sample 2,000 targets before retrieval (seed 20260712), with at most one target per paper, stratified by venue, year, and caption length.
The sample contains 732 NeurIPS, 706 ICLR, 386 ICML, and 176 TMLR targets, and 316 short, 573 medium, and 1,111 long captions under the sampling bins.

For each target, \texttt{gpt-5.6-sol} with medium reasoning effort writes a 35--90-word query using the four text fields.
An automatic validator checks that every target has exactly one query and rejects queries outside the length range or containing a figure number, the exact paper title, a copied six-token source span, a source-specific coined name, or a caption acronym.
Rejected queries are rewritten and checked again, yielding 2,000 accepted queries.
All experimental conditions use this fixed query set.
Because the queries express information from all four fields, the experiment evaluates how candidate text fields recover a known target under a controlled retrieval setting.

\subsection{Scoring and Metrics}

We encode each query and each candidate text field separately with \texttt{text-embedding-3-large} (3,072 dimensions).
The saved title, abstract, and caption vectors are matched to the release records, and citing paragraphs and queries use the same embedding model.
Vectors are L2-normalized, and similarities are computed in float32.
For query $q$, candidate $d$, and selected field set $S$, the retrieval score is
\begin{equation}
\mathrm{score}(q,d,S)=\frac{1}{|S|}\sum_{f\in S}
\cos\bigl(q,\mathrm{field}_{f}(d)\bigr).
\end{equation}
A retrieval script ranks all 52,052 candidates by this score.
The queries, candidate pool, embeddings, normalization, and scoring function are fixed; only the selected candidate fields change.

Each query has one target diagram.
R@$k$ is the percentage of queries whose target appears in the top $k$ results.
MRR@10 averages $1/r$ when the target has rank $r\leq10$ and zero otherwise, then multiplies by 100.
Higher scores are better for both metrics.
The primary comparison is the R@10 difference between the full field set and captions alone on the same 2,000 targets.

\subsection{Full Field Ablation}

Table~\ref{tab:retrieval_full} reports the full ablation.
Captions provide a strong retrieval signal, and citing paragraphs increase R@10 when added to the title, abstract, and caption.
The full field set retrieves 96 targets missed without citing paragraphs, while the reverse occurs for 32 targets.
The highest R@10 is obtained by combining captions, abstracts, and citing paragraphs.
This identifies a useful combination for this equal-weight scoring method.

\begin{table}[t]
\centering
\footnotesize
\setlength{\tabcolsep}{3pt}
\begin{tabularx}{\columnwidth}{@{}Lrrrr@{}}
\toprule
\textbf{Fields} & \textbf{R@1} & \textbf{R@5} & \textbf{R@10} & \textbf{MRR@10} \\
\midrule
C & \textbf{72.90} & 83.55 & 86.60 & 77.69 \\
C + T & 67.60 & 82.20 & 86.80 & 74.09 \\
C + T + A & 64.95 & 81.40 & 85.75 & 72.00 \\
Full & 68.60 & 84.90 & 88.95 & 75.79 \\
\midrule
Full without T & 71.75 & \textbf{87.35} & \textbf{90.75} & \textbf{78.48} \\
Full without A & 69.90 & 85.70 & 89.35 & 76.70 \\
Full without C & 48.85 & 69.95 & 76.10 & 58.12 \\
\midrule
Full, wrong P & 56.60 & 75.15 & 80.80 & 64.56 \\
Full, wrong T + A & 22.85 & 44.50 & 53.80 & 31.98 \\
\bottomrule
\end{tabularx}
\caption{Retrieval results on 2,000 queries and 52,052 candidates (\%). C: caption; T: title; A: abstract; P: citing paragraphs. Full uses all four fields; Full without P equals C + T + A. Wrong fields come from another paper matched on venue, year, and length bin. Bold marks the best score in each column.}
\label{tab:retrieval_full}
\end{table}

Captions achieve the highest R@1, while combining captions, abstracts, and citing paragraphs achieves the highest R@10 and MRR@10.

\subsection{Controls for Correct Pairing}

\noindent\textbf{Different-paper control.}
We replace a candidate's citing paragraphs with a passage from another paper matched by venue, year, and text length.
The same procedure is applied to title and abstract in a separate control.
These replacements hold the number of scored fields fixed while changing whether their contents match the candidate diagram.
Replacing citing paragraphs reduces R@10 from 88.95 to 80.80; replacing title and abstract reduces it to 53.80 (Table~\ref{tab:retrieval_full}).

\noindent\textbf{Same-paper control.}
We additionally swap citing paragraphs between diagrams from the same paper while keeping titles, abstracts, captions, and the candidate pool fixed.
A seeded procedure (20260929) forms disjoint pairs with different paragraph hashes, choosing an unused donor with the nearest log word count.
Records without a partner retain their original paragraphs.
This swaps the paragraph fields of 35,026 candidates and makes 913 of the 2,000 target queries eligible.
All conditions in Table~\ref{tab:within_paper_context}(b) use these same 913 queries.

Correctly paired paragraphs improve R@10 over the same-paper swap by 6.46 points.
The full field set exceeds captions alone by 2.63 points, while captions alone retain the highest R@1 and MRR@10 on this subset.
The control tests whether stored citing paragraphs distinguish diagrams within a paper; different paragraph fields can still share passages.

\FloatBarrier

\section{Diagram Generation}
\label{app:generation_protocols}

\subsection{PaperBananaBench Setup}

We use PaperBanana's implementation and original prompt templates (commit \texttt{e088a8f})~\cite{zhu2026paperbananaautomatingacademicillustration}.
Qwen3.8~\cite{qwen2026qwen38} serves as the planner and stylist, and Qwen-Image-2.1~\cite{wu2025qwenimage,qwen2026qwenimage21} renders each diagram with 40 sampling steps.
The four conditions use no references or ten references from PaperBanana's curated pool, Paper2Fig100k, or DiagramBank.
All reference conditions use Qwen3-Embedding-8B~\cite{zhang2025qwen3embedding} to retrieve 50 candidates and Qwen3.8 to rerank them, selecting ten references with at most one diagram per paper.

\noindent\textbf{Reference pool.}
The initial DiagramBank pool contains 37,754 DiagramBank-Core records from 19,558 papers, selected by citing-paragraph availability and the source-record date filter.
The exact filter is recorded in the accompanying \texttt{generation\_reference\_protocol.json} configuration.
We remove 541 papers with titles identical or near-identical to PaperBananaBench test papers or to papers in PaperBanana's reference pool.
For each test case, we also exclude papers sharing an author with the target and records sharing a 13-word span with its caption or method section.
The target diagram is supplied only to the evaluator.

\noindent\textbf{Reference inputs.}
Each DiagramBank reference supplies its image, resized to a maximum longer-side length of 1,024 pixels, together with its abstract (up to 250 words) and citing paragraphs (up to 600 words).
References occupy the in-context example positions of the original PaperBanana prompts.
The mean planner prompt contains 14.4k tokens with DiagramBank references and 43.5k with PaperBanana references.

\subsection{Evaluation and Results}

\noindent\textbf{PaperBanana score.}
Qwen3.8 compares each generated diagram with the target diagram on faithfulness, conciseness, readability, and aesthetics.
A generated-diagram win receives 100, a tie 50, and a loss 0.
The overall verdict follows PaperBanana's hierarchical aggregation: faithfulness and readability are considered first, followed by conciseness and aesthetics when the first pair does not yield a winner~\cite{zhu2026paperbananaautomatingacademicillustration}.
At either level, a win on both criteria, or one win and one tie, determines the winner.
Scores are averaged over the 292 cases; the overall PaperBanana score is therefore not the arithmetic mean of the four dimension scores.
Table~\ref{tab:pbb} reports the dimension scores and gains.

\noindent\textbf{Direct pairwise evaluation.}
A separate rubric presents the target diagram and two generated diagrams and compares faithfulness, text, structure, and visual quality in both presentation orders.
Scores lie in $[-1,1]$, with positive values favoring the named reference pool over its comparison condition.
Qwen3.8 judges DiagramBank against no references and PaperBanana references on all 292 cases, while Claude Haiku~4.5 judges DiagramBank against no references on the same cases.
The Paper2Fig100k condition uses the same retriever and filters and is compared with no references under the Qwen3.8 rubric on all 292 cases.
Table~\ref{tab:pbb_rubric_details} reports the criterion scores, overall pairwise preference score, and case-level wins, ties, and losses.

\begin{table*}[t]
\centering
\small
\setlength{\tabcolsep}{5pt}
\begin{tabular*}{\textwidth}{@{\extracolsep{\fill}}llrrrrrr@{}}
\toprule
\textbf{Compared with} & \textbf{Judge} & \textbf{Faith.} & \textbf{Text} & \textbf{Struct.} & \textbf{Visual} & \textbf{Overall} & \textbf{W/T/L} \\
\midrule
\multicolumn{8}{@{}l}{\textit{(a) DiagramBank references compared with}} \\
PaperBanana references & Qwen3.8 & +0.14 & +0.16 & +0.20 & +0.22 & +0.188 & 151/45/96 \\
No references & Qwen3.8 & +0.39 & +0.35 & +0.36 & +0.37 & +0.404 & 190/30/72 \\
No references & Haiku 4.5 & +0.25 & +0.28 & +0.31 & +0.25 & +0.339 & 139/113/40 \\
\midrule
\multicolumn{8}{@{}l}{\textit{(b) Paper2Fig100k references compared with}} \\
No references & Qwen3.8 & +0.17 & +0.15 & +0.14 & +0.16 & +0.18 & 156/32/104 \\
\bottomrule
\end{tabular*}
\caption{Pairwise preference scores on 292 PaperBananaBench cases. Scores are in $[-1,1]$; positive values favor the reference pool named in each block. W/T/L counts its wins, ties, and losses. DiagramBank and Paper2Fig100k use the same no-reference baseline. Table~\ref{tab:pbb_rubric} summarizes the Qwen3.8 pairwise preference scores.}
\label{tab:pbb_rubric_details}
\end{table*}

\subsection{Additional Judges and Renderers}

\begin{table}[!htbp]
\centering
\small
\setlength{\tabcolsep}{6pt}
\begin{tabular*}{\columnwidth}{@{\extracolsep{\fill}}lrr@{}}
\toprule
\multicolumn{3}{@{}l}{\textit{(a) Judges: pairwise preference score}} \\
\textbf{Judge} & \textbf{Overall} & \textbf{W/T/L} \\
\midrule
Qwen3.8 & +0.708 & 39/4/5 \\
Sonnet 5.5 & +0.708 & 40/2/6 \\
\midrule
\multicolumn{3}{@{}l}{\textit{(b) Renderers: PaperBanana score}} \\
\textbf{Renderer} & \textbf{No refs.} & \textbf{DiagramBank} \\
\midrule
Qwen-Image-2.1 & 0.00 & 10.42 \\
Sonnet 5.5 (SVG) & 56.3 & 59.4 \\
\bottomrule
\end{tabular*}
\caption{Additional judges and renderers on 48 shared cases. (a) Pairwise preference scores in $[-1,1]$; positive values favor DiagramBank over no references, with W/T/L giving its wins, ties, and losses. (b) PaperBanana scores (0--100), using fixed rendering prompts and Qwen3.8 as judge.}
\label{tab:pbb_calibration}
\label{tab:renderer_check}
\end{table}

On the same 48-case subset, Claude Sonnet~5.5 and Qwen3.8 both favor DiagramBank references over no references (Table~\ref{tab:pbb_calibration}(a)).
Sonnet's criterion scores are also positive: faithfulness $+0.67$, text $+0.62$, structure $+0.28$, and visual quality $+0.17$.

\begin{figure*}[t]
\centering
\begin{subfigure}[t]{0.80\textwidth}
\centering
\includegraphics[width=\linewidth]{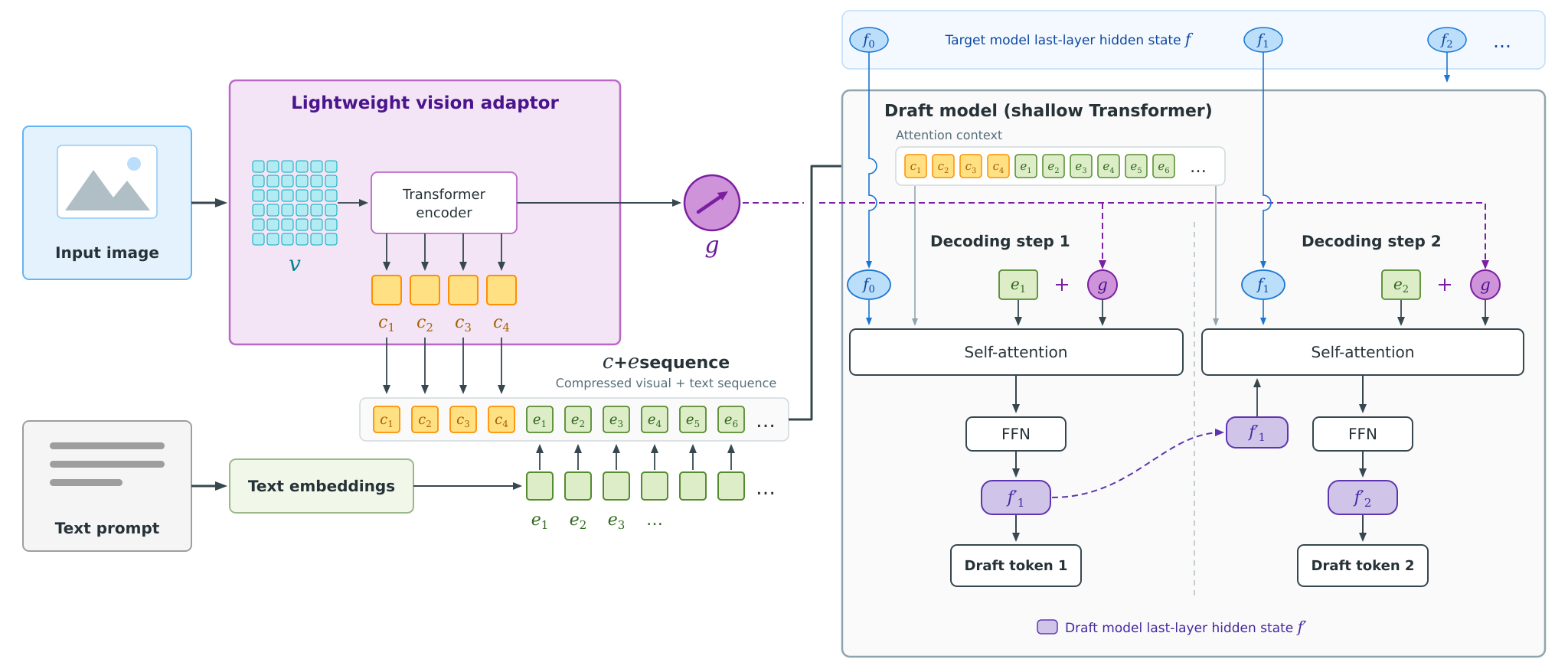}
\caption{No references}
\label{fig:svg_example_c0}
\end{subfigure}

\medskip
\begin{subfigure}[t]{0.80\textwidth}
\centering
\includegraphics[width=\linewidth]{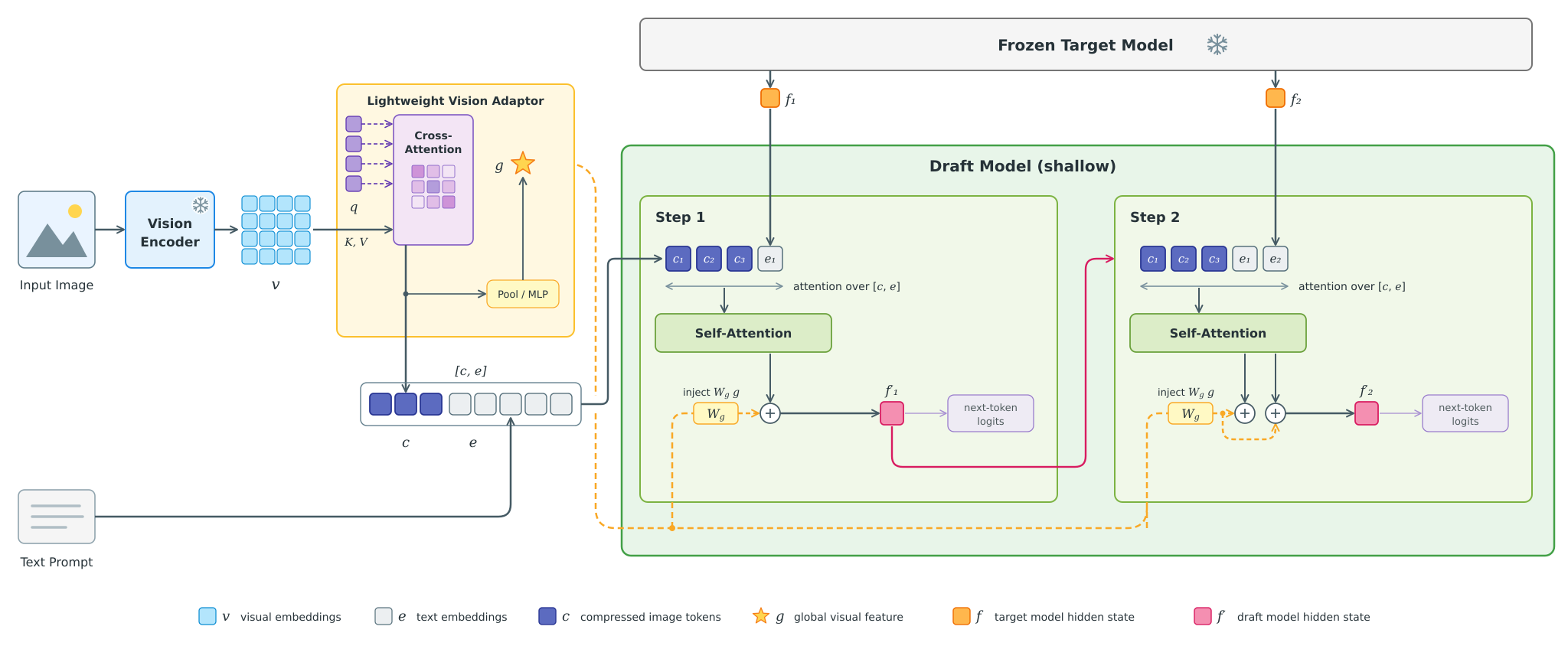}
\caption{DiagramBank references}
\label{fig:svg_example_c2}
\end{subfigure}
\caption{SVG diagrams for the ViSpec case in PaperBananaBench. Both conditions use the same Qwen3.8 planner and stylist with Sonnet 5.5 as the SVG renderer. With DiagramBank references, the output includes the frozen vision encoder, separates decoding steps, and adds a legend for the notation. The output without references omits the vision encoder.}
\label{fig:svg_example}
\end{figure*}

For the renderer comparison, we reuse the rendering prompts produced with and without DiagramBank references for these 48 cases and replace Qwen-Image-2.1 with Claude Sonnet~5.5 generating SVG code.
Qwen3.8 assigns PaperBanana scores to outputs from both renderers.
DiagramBank improves the overall PaperBanana score for both renderers (Table~\ref{tab:renderer_check}(b)).
Figure~\ref{fig:svg_example} illustrates the SVG outputs for one case.

\subsection{BizGenEval}
\label{app:bizgeneval}

BizGenEval contains 400 prompts in five domains: scientific figures, slides, webpages, posters, and charts, with 80 prompts per domain~\cite{li2026bizgeneval}.
Each prompt has 20 binary checklist questions answered by Qwen3.8.
We report question accuracy (Acc.) and the benchmark's Easy, Hard, and All scores.
The generation pipeline is fixed across no references, ten random DiagramBank references, and ten retrieved DiagramBank references.

Table~\ref{tab:bizgeneval} reports the scientific-figure results, and Table~\ref{tab:bizgeneval_paired}(a,b) gives the paired gains and case outcomes.
Retrieved references improve all four metrics over both baselines.
Table~\ref{tab:bizgeneval_other}(c) extends the comparison to the remaining four domains, showing that the benefit of reference selection varies by domain.

\begin{table}[!htbp]
\centering
\small
\setlength{\tabcolsep}{6pt}
\begin{tabular*}{\columnwidth}{@{\extracolsep{\fill}}lrr@{}}
\toprule
\textbf{Metric} & \textbf{Gain} & \textbf{W/T/L} \\
\midrule
\multicolumn{3}{@{}l}{\textit{(a) Retrieved vs.\ no references}} \\
Acc. & +2.06 & 39/13/28 \\
Easy & +5.50 & 27/37/16 \\
Hard & +0.50 & 24/32/24 \\
All & +4.50 & 35/21/24 \\
\midrule
\multicolumn{3}{@{}l}{\textit{(b) Retrieved vs.\ random references}} \\
Acc. & +1.81 & 39/14/27 \\
Easy & +2.75 & 26/34/20 \\
Hard & +1.75 & 28/33/19 \\
All & +2.25 & 35/20/25 \\
\midrule
\multicolumn{3}{@{}l}{\textit{(c) Other domains vs.\ no references}} \\
\textbf{References} & $\boldsymbol{\Delta}$\textbf{Acc.} & $\boldsymbol{\Delta}$\textbf{All} \\
\midrule
Random & +2.28 & +4.00 \\
Retrieved & +0.42 & +0.16 \\
\bottomrule
\end{tabular*}
\caption{BizGenEval gains. (a,b) Paired comparisons on 80 scientific-figure prompts; W/T/L counts higher, equal, and lower scores. (c) Gains on the other 320 prompts. Each reference condition uses ten DiagramBank images; gains are computed before rounding.}
\label{tab:bizgeneval_paired}
\label{tab:bizgeneval_other}
\end{table}

\subsection{Blind Human Evaluation}
\label{app:human_evaluation}

\begin{table}[!htbp]
\centering
\small
\setlength{\tabcolsep}{7pt}
\begin{tabular*}{\columnwidth}{@{\extracolsep{\fill}}lrrr@{}}
\toprule
 & \multicolumn{3}{c@{}}{\textbf{Preferred condition}} \\
\cmidrule(l){2-4}
\textbf{Evaluator} & \textbf{DB} & \textbf{None} & \textbf{Tie} \\
\midrule
1 & 13 & 5 & 2 \\
2 & 17 & 2 & 1 \\
3 & 16 & 4 & 0 \\
4 & 13 & 3 & 4 \\
\midrule
Total votes (80) & 59 & 14 & 7 \\
Case outcomes (20) & 14 & 1 & 5 \\
\bottomrule
\end{tabular*}
\caption{Blind human evaluation of outputs generated with DiagramBank references (DB) and without references (None). Each evaluator judges 20 pairs; a case win requires at least three of four votes.}
\label{tab:generation_study}
\end{table}

\noindent\textbf{Cases and generation.}
We select 20 DiagramBank records before generation, with five cases per venue and fixed caption-length quotas.
Each scientific brief contains the title, abstract, caption, and citing paragraphs of its target record.
References are the top three candidates under the full-field retrieval score in Appendix~\ref{app:retrieval_protocols}, excluding the target paper and identifiers with the same normalized title and abstract.
All 60 reference images pass decoding, SHA-256 comparison, and a 64-bit DCT perceptual-hash check against the target and each other, using Hamming distance 5 as the near-duplicate threshold.

Each brief is rendered once without references and once with three reference images, using identical text prompts.
The prompt requests a faithful scientific diagram with readable labels and prohibits adding unsupported components or copying scientific content from the reference images.
We use Codex's \texttt{image\_gen} tool in separate sessions.

\noindent\textbf{Evaluation protocol.}
Four evaluators independently compare all 20 pairs using the supplied scientific brief.
They judge overall preference, prioritizing faithful components and relationships, readable labels, and clear visual organization.
Case order is randomized separately for each evaluator.
Left/right positions are counterbalanced, with DiagramBank appearing on each side for ten cases per evaluator; condition names, filenames, retrieval scores, and source identifiers are hidden.
Evaluators choose either diagram or a tie.
A case is a win only when at least three of four evaluators select the same condition; all other vote patterns count as ties.

\noindent\textbf{Results.}
Table~\ref{tab:generation_study} provides the individual votes underlying the summary in Table~\ref{tab:human_summary}.
Fleiss' $\kappa$ across the three response categories is 0.222.

\FloatBarrier

\end{document}